\documentclass[a4paper]{revtex4}
\usepackage{graphicx}
\usepackage{fancyhdr}
\usepackage{amsmath}
\usepackage{color}

\begin{document}

\title{Multiple Point Principle of the Gauged B-L Model}

%

\author{K. Kawana}
\affiliation{Department of Physics, Kyoto University, Kyoto 606-8502, Japan}

\begin{abstract}
We consider the multiple point principe (MPP), its breaking and the inflation of the gauged B-L model. Our results are based on the two-loop renormalization group equations (RGEs). We find that this model can realize the flat potentials at $\Lambda_{\text{MPP}}=10^{17}$GeV, and that the unknown parameters of the model are fixed by the MPP. Furthermore, we show that it is possible to realize both of the electroweak symmetry breaking at ${\cal{O}}(100)$GeV and the inflation by breaking the MPP.
\end{abstract}

\maketitle

\thispagestyle{fancy}


\section{Introduction}
The discovery of the Higgs like particle and its mass \cite{Aad:2012tfa,Chatrchyan:2012ufa} is very meaningful for the Standard Model (SM). The experimental value of the Higgs mass suggests that the Higgs potential can be stable up to the Planck scale $M_{pl}$ and also that both of the Higgs self coupling $\lambda$ and its beta function $\beta_{\lambda}$ become very small around $M_{pl}$. This fact attracts much attention, and there are many works which try to find its physical meaning \cite{Froggatt:1995rt,Froggatt:2001pa,Nielsen:2012pu,Buttazzo:2013uya,Shaposhnikov:2009pv,Meissner:2007xv,Khoze:2014xha,Kawai:2011qb,Kawai:2013wwa,Hamada:2014ofa,Hamada:2014xra,Bezrukov:2007ep,Hamada:2013mya,Hamada:2014iga,Hamada:2014wna,Hamada:2014raa,Kawamura:2013kua,Meissner:2006zh,Haba:2014sia,Hamada:2015ria}.

Well before the discovery of the Higgs, it was argued that the Higgs mass can be predicted to be around $130$GeV by the requirement that the Higgs potential becomes flat at $M_{pl}$ \cite{Froggatt:1995rt,Froggatt:2001pa}. Such a requirement (not always at $M_{pl}$) is generally called the multiple point principle (MPP). One of the good points of the MPP is its predictability to the parameters of a low energy effective  theory. For example, the values of the scalar couplings are radiatively generated by the renormalization group equations (RGEs), and the flatness of the potentials furthermore gives the strong relations between parameters. As a result, the low-energy effective couplings can be uniquely predicted by the MPP. See \cite{Hamada:2014xka,Kawana:2014zxa} for example.

It is meaningful to consider whether such a criticality can be also realized in the models beyond the SM. One of the phenomenologically interesting models is the minimal gauged B-L model \cite{Iso:2009ss,Iso:2012jn,Okada:2010wd,Okada:2011en,Kawana:2015tka}. In \cite{Iso:2012jn}, it was argued that this model can naturally explain the electroweak scale by applying the MPP to the Higgs potential. In this paper, we study the MPP of the whole scalar potential and its implications based on the two-loop renormalization equations (RGEs). We also consider the inflation where the newly introduced complex scalar $\Psi$ plays the roll of the inlfaton. Here, we do not follow all the calculations. See \cite{Kawana:2015tka} for the details.

\section{MPP of the Gauged B-L Model}
\begin{figure}
\begin{center}
\begin{tabular}{c}
\begin{minipage}{0.5\hsize}
\begin{center}
\includegraphics[width=8.5cm]{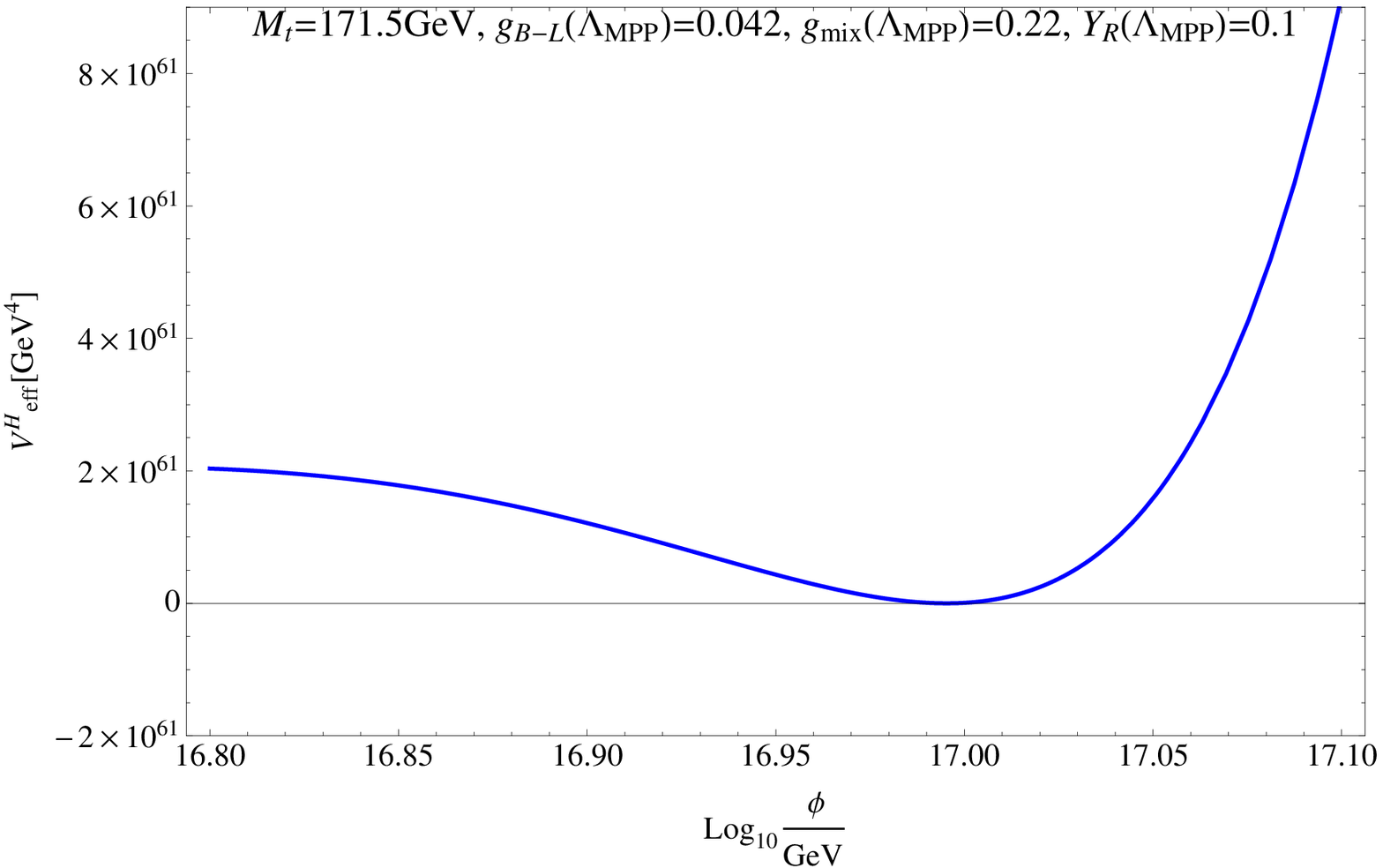}
\end{center}
\end{minipage}
\begin{minipage}{0.5\hsize}
\begin{center}
\includegraphics[width=8.5cm]{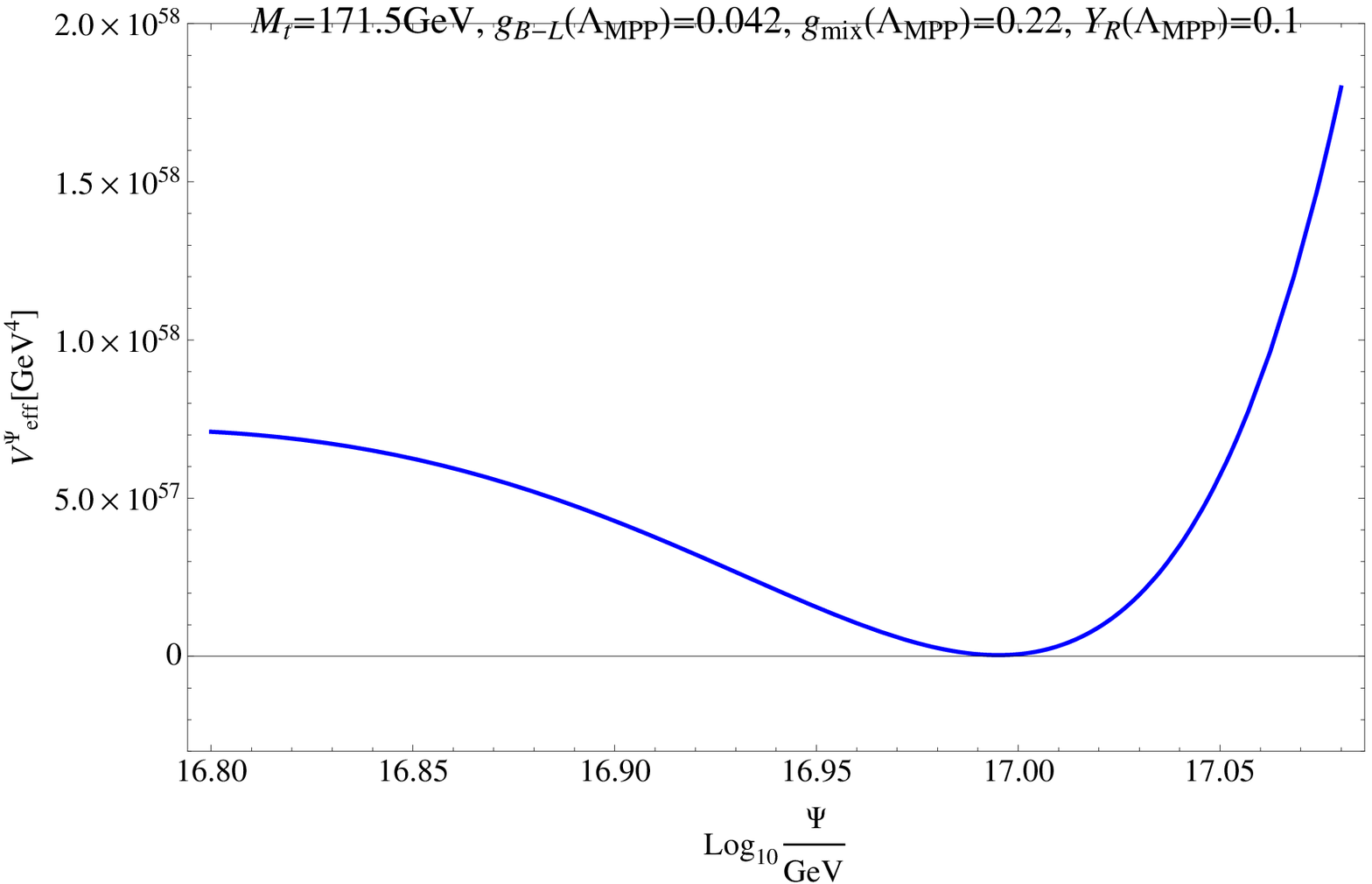}
\end{center}
\end{minipage}
\\
\\
\begin{minipage}{0.5\hsize}
\begin{center}
\includegraphics[width=8.5cm]{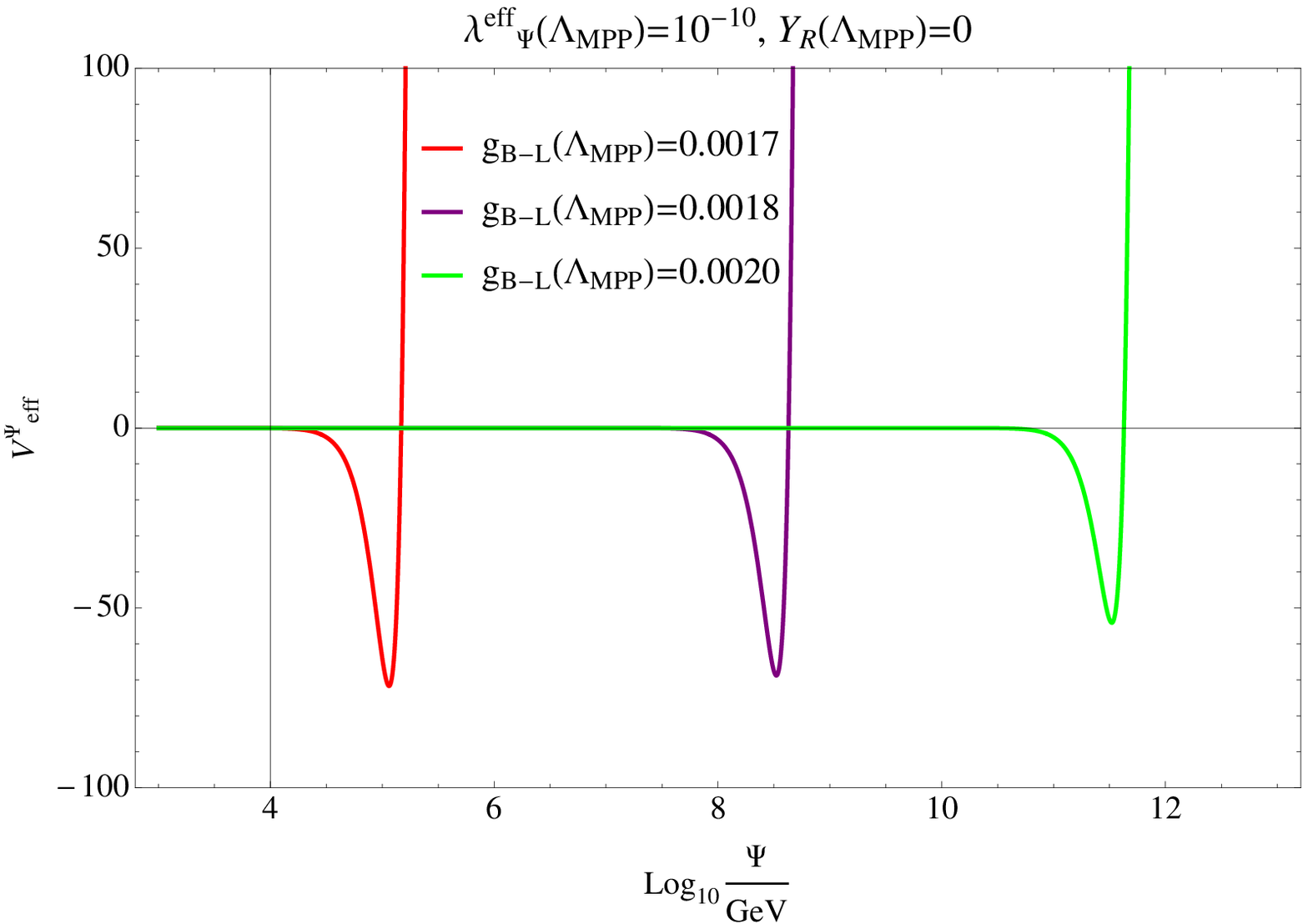}
\end{center}
\end{minipage}
\begin{minipage}{0.5\hsize}
\begin{center}
\includegraphics[width=8.5cm]{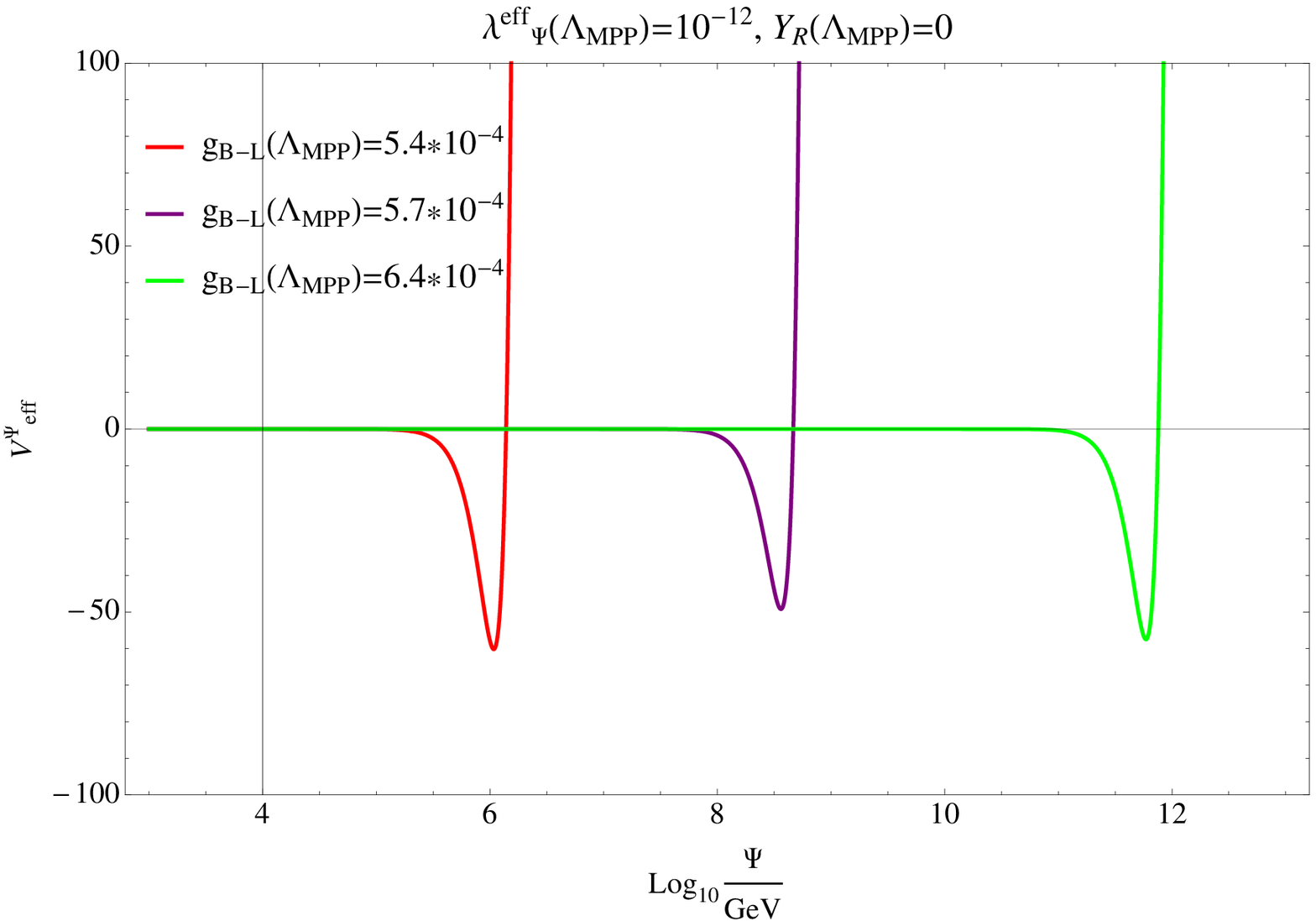}
\end{center}
\end{minipage}
\end{tabular}
\end{center}
\caption{The upper left (right) panel shows the effective potential $V_{\text{eff}}^{H}\hspace{1mm}(V_{\text{eff}}^{\Psi})$ when the MPP conditions are satisfied. They are exactly flat at $\Lambda_{\text{MPP}}=10^{17}$GeV. On the other hand, the lower panels show $V_{\text{eff}}^{\Psi}$ when the MPP is broken. The left (right) panel corresponds to $\lambda_{\Psi}(\Lambda_{\text{MPP}})=10^{-10}\hspace{1mm}(10^{-12})$.}
\label{fig:1}
\end{figure}

In addition to $\lambda$, there are two scalar couplings in this model: 
\begin{equation} {\cal{L}}\ni-\lambda_{\Psi}\left(\Psi^{\dagger}\Psi\right) -\kappa \left(H^{\dagger}H\right)\left(\Psi^{\dagger}\Psi\right). \end{equation} 
Therefore, the whole MPP conditions are
\begin{equation} \lambda(\Lambda_{\text{MPP}})=\lambda_{\Psi}(\Lambda_{\text{MPP}})=\kappa(\Lambda_{\text{MPP}})=\beta_{\lambda}(\Lambda_{\text{MPP}})=\beta_{\lambda_{\Psi}}(\Lambda_{\text{MPP}})=\beta_{\kappa}(\Lambda_{\text{MPP}})=0,\label{eq:con1}\end{equation}
where $\Lambda_{\text{MPP}}$ is a high energy scale at which we impose the MPP. In this paper, $\Lambda_{\text{MPP}}$ is chosen to be $10^{17}$GeV. In \cite{Kawana:2015tka}, it is shown that we can actually realize Eq.(\ref{eq:con1}) simultaneously and that the unknown parameters of the model are uniquely determined by these conditions.  For example, the top mass $M_{t}$ and the B-L gauge coupling $g_{B-L}$ are predicted to be 
\begin{equation} M_{t}\simeq 176\text{GeV}\hspace{2mm},\hspace{2mm}g_{B-L}(\Lambda_{\text{MPP}})\simeq0.\end{equation}
The upper panels of Fig.\ref{fig:1} show the effective potentials when Eq.(\ref{eq:con1}) is satisfied. One can see that they are exactly flat at $10^{17}$GeV.

\section{Breaking of MPP}
As discussed in \cite{Iso:2012jn,Kawana:2015tka}, if we want to realize the electroweak symmetry breaking by the radiative $U(1)_\text{{B-L}}$ breaking, we need to break the MPP.
Namely, $\lambda_{\Psi}$ and its beta function $\beta_{\lambda_{\Psi}}$ should be positive at $\Lambda_{\text{MPP}}$ to realize the Coleman-Weinberg mechanism. The lower panels in Fig.\ref{fig:1} show such examples. Here, we choose $\lambda_{\Psi}(\Lambda_{\text{MPP}})=10^{-10}(10^{-12})$ in the left (right) panel. The different colors corresponds to the different values of $g_{B-L}(\Lambda_{\text{MPP}})$. 

After the $U(1)_\text{{B-L}}$ symmetry breaking, the Higgs mass term is generated by 
\begin{equation} {\cal{L}}\ni -\kappa \left(H^{\dagger}H\right)\left(\Psi^{\dagger}\Psi\right).\end{equation}
As a result, the Higgs expectation value $v_{h}$ is given by
\begin{equation}v_{h}=\sqrt{-\frac{\kappa}{\lambda}}\times v_{B-L},\label{eq:vev}\end{equation}
where $v_{B-L}$ is the expectation value of $\Psi$. In \cite{Kawana:2015tka}, it is shown that we can realize $v_{h}={\cal{O}}(100)$GeV by tuning the parameters of the model. For example, the red contour of the lower left panel in Fig.\ref{fig:1} actually corresponds to such case. Here, $v_{B-L}={\cal{O}}(10^{5})$GeV and $\kappa(v_{h})={\cal{O}}(10^{-7})$. Thus, the right hand side of Eq.(\ref{eq:vev}) becomes ${\cal{O}}(100)$GeV.

\section{Inflation}
\begin{figure}
\begin{center}
\begin{tabular}{c}
\begin{minipage}{0.5\hsize}
\begin{center}
\includegraphics[width=8cm]{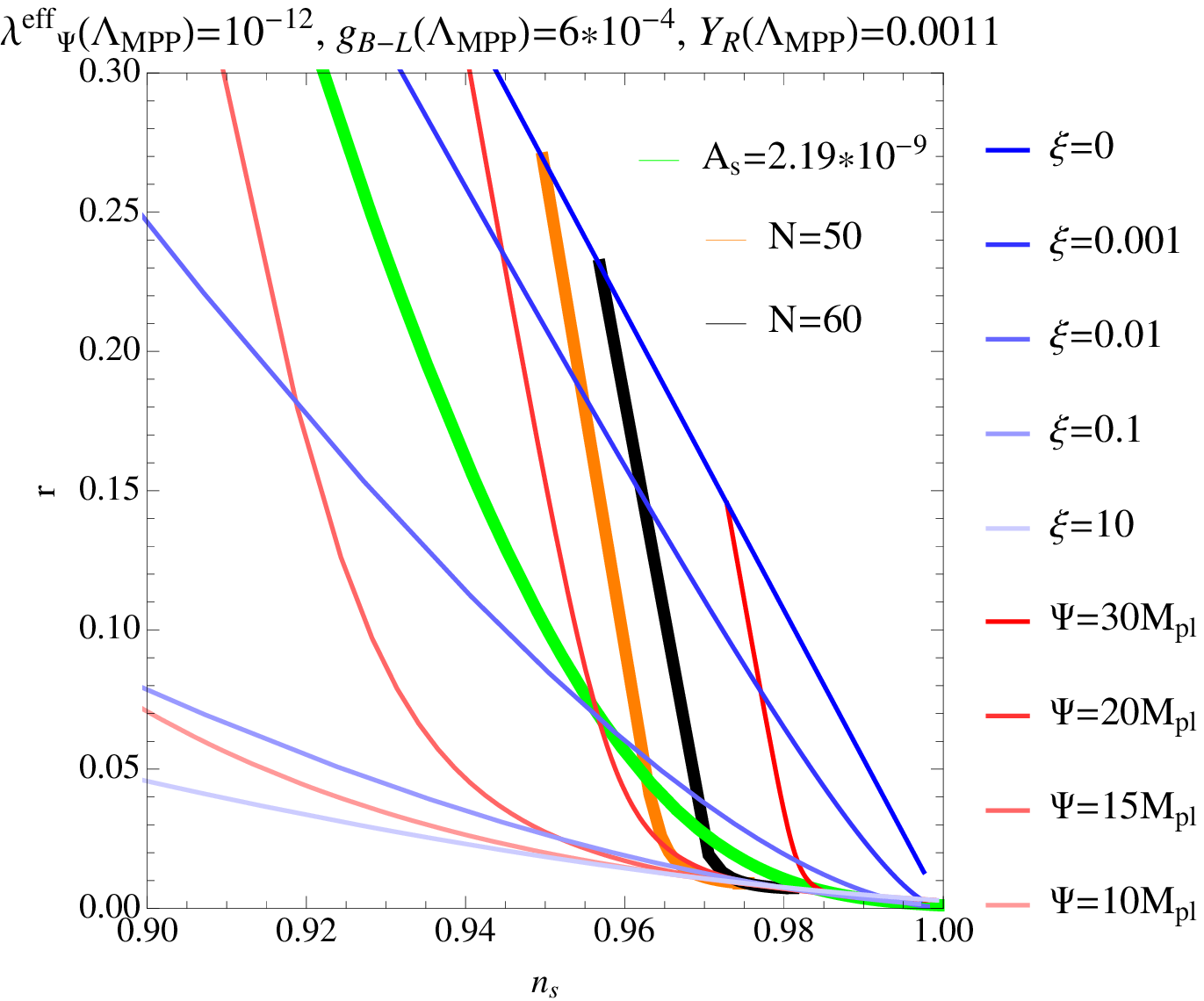}
\end{center}
\end{minipage}
\begin{minipage}{0.5\hsize}
\begin{center}
\includegraphics[width=8cm]{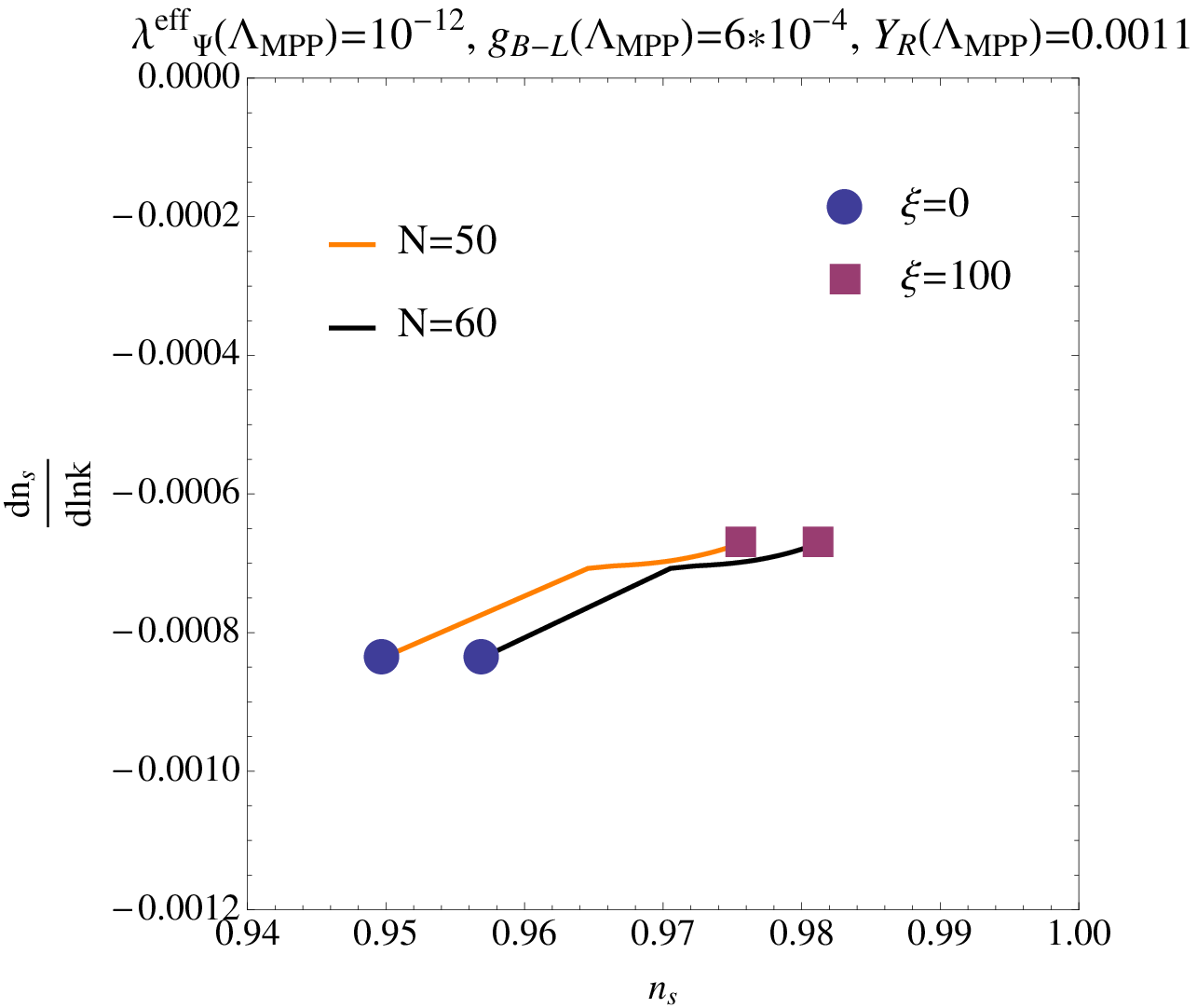}
\end{center}
\end{minipage}
\end{tabular}
\end{center}
\caption{The cosmological predictions of the gauged B-L model. Here, $\lambda_{\Psi}(\Lambda_{\text{MPP}})$ is chosen to be $10^{-12}$. The left (right) panel shows $n_{s}$ vs $r \hspace{1mm}(dn_{s}/\ln k)$.}
\label{fig:2}
\end{figure}

As well as the Higgs inflation with the non-minimal gravitational coupling $\xi {\cal{R}} \phi^{2}$, we can also consider the inflation scenario where $\Psi$ plays a roll of the inflaton. Here, we show the cosmological predictions of this scenario. The left (right) panel in Fig.\ref{fig:2} shows the spectral index $n_{s}$ vs the scalar-to-tensor ratio $r$ (its running $dn_{s}/d\ln k$). Here, $\lambda_{\Psi}(\Lambda_{\text{MPP}})$ is chosen to be $10^{-12}$. In the left panel, the green contour corresponds to the observed CMB fluctuation \cite{Ade:2013zuv}
\begin{equation} A_{s}:=\frac{V_{\Psi}}{24\pi^{2}\epsilon M_{pl}^{4}}=\left(2.196^{+0.053}_{-0.058}\right)\times10^{-9}\hspace{2mm}(\text{68\% CL}),\end{equation}
and the orange (black) contour corresponds to the number of e-foldings $N=50$ (60). In the right panel, we change the non-minimal coupling $\xi$ from 0 to 100. These results are consistent with the current cosmological constraints by Planck+WMAP \cite{Ade:2013zuv} except for the small values of $dn_{s}/\ln k$. We might improve this situation by considering a higher dimensional operator \cite{Hamada:2014wna}. 

\section{Summary}
We have considered the MPP and the inflation of the gauged B-L model. We have found that the potentials of the scalar fields can become flat at $\Lambda_{\text{MPP}}=10^{17}$GeV, and that the parameters of the models can be uniquely fixed by the MPP. We have also seen that it is possible to realize the electroweak symmetry breaking at ${\cal{O}}(100)$GeV by breaking the MPP. The inflation scenario where $\Psi$ plays a roll of the inflaton is consistent with the current cosmological constraints. 
\begin{acknowledgments}
K.Kawana thanks organizers of ``{\it {
		Toyama International Workshop on Higgs as a Probe of New Physics 2015 }}'' for hospitality.
\end{acknowledgments}

\bigskip 

\end{document}